\documentclass[10pt,twoside,twocolumn,english,aps,prl,superscriptaddress,notitlepage]{revtex4-2}
\usepackage[utf8]{inputenc}
\usepackage[letterpaper]{geometry}
\usepackage{color,xcolor,soul,multirow}
\usepackage{verbatim,textcomp}
\usepackage{amstext,amsfonts,amsmath,amssymb}
\usepackage{graphicx}
\usepackage[calcwidth,explicit]{titlesec}
\usepackage{chngcntr}
\usepackage[normalem]{ulem}
\usepackage[unicode=true,pdfusetitle, bookmarks=false, breaklinks=true,pdfborder={0 0 0},pdfborderstyle={},backref=false,colorlinks=true]{hyperref}
\hypersetup{colorlinks=true,citecolor=blue,linkcolor=black,urlcolor=blue}

\makeatletter\renewcommand\frontmatter@abstractwidth{\dimexpr\textwidth-3cm\relax}\makeatother

\titleformat{\section}{}{}{0pt}{}
\titleformat{\section}{\bfseries\sffamily\large\filcenter}{\thesection.}{0.2em}{#1}
\titlespacing{\section}{0pt}{0.2ex}{0.2ex}
\titleformat{\paragraph}[runin]{\normalfont\normalsize\bfseries}{}{0pt}{\theparagraph.}
\titlespacing*{\paragraph}{0em}{0ex}{0.3em}[]

{}

\renewcommand\thesection{\Roman{section}}
\counterwithout{paragraph}{subsubsection}
\renewcommand{\theparagraph}{\arabic{paragraph}}
\makeatletter\def\p@paragraph{}\makeatother

\AtBeginDocument{
\renewcommand{\ref}[1]{\autoref{#1}}
\renewcommand{\figureautorefname}{Fig.}\renewcommand{\equationautorefname}{Eq.}
\renewcommand{\tableautorefname}{Tbl.}
\renewcommand{\sectionautorefname}{\S}\renewcommand{\subsectionautorefname}{\S}
\renewcommand{\paragraphautorefname}{\S}
}

\begin{document}
\setlength{\abovedisplayskip}{0.2ex}\setlength{\belowdisplayskip}{0.2ex}
\setlength{\abovedisplayshortskip}{0.2ex}\setlength{\belowdisplayshortskip}{0.2ex}

\title{Thermal evolution of skyrmion formation mechanism in chiral multilayer films}

\author{Xiaoye Chen}\thanks{These authors contributed equally to this work}
\email{chen\_xiaoye@imre.a-star.edu.sg}
\affiliation{Institute of Materials Research \& Engineering, Agency for Science, Technology \& Research (A{*}STAR), 138634 Singapore}
\author{Edwin Chue}\thanks{These authors contributed equally to this work}
\affiliation{Physics Department, National University of Singapore (NUS), 117551
Singapore}

\author{Jian Feng Kong}
\affiliation{Institute of High Performance Computing, Agency for Science,
Technology \& Research (A{*}STAR), 138632 Singapore}
\author{Hui Ru Tan}
\affiliation{Institute of Materials Research \& Engineering, Agency for Science, Technology \& Research (A{*}STAR), 138634 Singapore}
\author{Hang Khume Tan}
\affiliation{Institute of Materials Research \& Engineering, Agency for Science, Technology \& Research (A{*}STAR), 138634 Singapore}

\author{Anjan Soumyanarayanan}
\email{anjan@nus.edu.sg}
\affiliation{Physics Department, National University of Singapore (NUS), 117551 Singapore}
\affiliation{Institute of Materials Research \& Engineering, Agency for Science, Technology \& Research (A{*}STAR), 138634 Singapore}

\begin{abstract}
\noindent 
Magnetic skyrmions form in chiral multilayers from the shrinking or fission of elongated stripe textures. Here we report an experimental and theoretical study of the temperature dependence of this stripe-to-skyrmion transition in Co/Pt-based multilayers. Field-reversal magnetometry and Lorentz microscopy experiments over 100 -- 350~K establish the increased efficacy of stripe-to-skyrmion fission at higher temperatures --- driven primarily by the thermal evolution of key magnetic interactions --- thereby enhancing skyrmion density. Atomistic calculations elucidate that the energy barrier to fission governs the thermodynamics of the skyrmion formation. Our results establish a mechanistic picture of the stripe-to-skyrmion transition and advance the use of thermal knobs for efficient skyrmion generation.   
\end{abstract}
\maketitle

\section*{Introduction\label{sec:intro}}

Magnetic skyrmions are topological spin textures arising from the interplay of collinear and chiral magnetic interactions \cite{nagaosa_topological_2013, wiesendanger_nanoscale_2016, soumyanarayanan_emergent_2016}. They exhibit myriad technologically desirable characteristics within multilayer thin films such as ambient stability \cite{moreau-luchaire_additive_2016, boulle_room-temperature_2016}, tunable, nanoscale sizes \cite{soumyanarayanan_tunable_2017, romming_field-dependent_2015}, topological protection \cite{rosler_spontaneous_2006, hagemeister_stability_2015,zheng_targetsk_2017}, and efficient coupling to electrical currents \cite{romming_writing_2013, jiang_blowing_2015,woo_observation_2016}. These attributes have prompted explorations of their use as building blocks for next-generation computing \cite{fert_magnetic_2017, song_skyrmion-based_2020, zazvorka_thermal_2019}. One crucial prerequisite for most applications is the controlled generation of skyrmions within device configurations.

Several works have investigated skyrmion nucleation within a uniformly magnetized background --- using constrictions \cite{jiang_blowing_2015,finizio_deterministic_2019}, spatial defects \cite{buttner_field-free_2017,woo_deterministic_2018}, or localized heating \cite{je_creation_2018, je_targeted_2021}. Meanwhile, the zero-field (ZF) configuration for chiral multilayers typically comprises elongated stripes, which transform into skyrmions at moderate out-of-plane (OP) fields \cite{romming_writing_2013,leonov_properties_2016,moreau-luchaire_additive_2016,woo_observation_2016,soumyanarayanan_tunable_2017}. Recent works have shown that such stripes can also be transformed into metastable ZF skyrmions by applied electrical currents \cite{lemesh_current-induced_2018,ang_electrical_2020,brock_current-induced_2020,wang_electric-field-driven_2020}. While this \textbf\emph{{driven stripe-to-skyrmion transition}} is largely attributed to thermal excitations \cite{lemesh_current-induced_2018,brock_current-induced_2020,wang_electric-field-driven_2020}, the associated mechanism and energetics remain to be established. Moreover, the temperature dependence of this transition, crucial to its practical utility, remains unaddressed.

Skyrmion formation is governed by magnetic interactions describing domain wall (DW) energetics \cite{bogdanov_chiral_2001,rohart_skyrmion_2013}. For multilayers hosting the interfacial Dzyaloshinskii-Moriya interaction (iDMI), chiral domain stability can be described by the dimensionless parameter $\kappa$ \cite{bogdanov_chiral_2001, rohart_skyrmion_2013, soumyanarayanan_tunable_2017}, defined as $\kappa = \pi D / 4 \sqrt{AK_\text{eff}}$, where $D$ is the iDMI, $A$ is the direct exchange. The effective anisotropy, $K_\text{eff} \equiv K_{\rm u} - \mu_0 M_{\rm s}^2/2$ includes uniaxial ($K_{\rm u}$) and shape anisotropy ($-\mu_0 M_{\rm s}^2/2$) contributions, where $M_{\rm s}$ is the saturation magnetization. Recently, it has been shown that for chiral multilayers with $\kappa > 1$, skyrmions may be formed from the break-up, or ``fission'' of stripes \cite{tan_skyrmion_2020, chen_unveiling_2022}. Temperature may influence the transition energetics and kinetics, and therefore merits systematic investigation.

Here, we report on the temperature dependence of the stripe-to-skyrmion transition using thermodynamic and microscopic probes. Across four chiral multilayers over 100--350~K, using first-order reversal curve (FORC) magnetometry, we find that features associated with stripe fission become more prominent with increasing temperature. Concomitantly, Lorentz transmission electron microscopy (LTEM) measurements establish that each stripe fissions into more skyrmions, thereby enhancing the skyrmion density. These experimentally measured signs of the enhanced propensity of fission with increasing temperature are correlated with an increase in $\kappa$. With geodesic nudged elastic band (GNEB) calculations, we show that the observed temperature evolution of the stripe-to-skyrmion transition is ultimately due to the reduction in energy barriers to fission, thereby boosting the rate of this transition.

This work leverages [Ir(10)/Fe($x$)/Co($y$)/Pt(10)]$_{14}$ multilayer platform (thickness in angstroms in parentheses), wherein $D$ and $K_\text{eff}$ can be widely varied via the ratio of Fe and Co thicknesses \cite{soumyanarayanan_tunable_2017}. The four samples studied here --- sputter deposited on Si/SiO$_2$ substrates (see SM1) --- have 1~nm thick FM layers, and are referred to by their Fe($x$)/Co($y$) composition. They were shown to host sub-100~nm Néel-textured skyrmions whose RT structure and stability have been studied extensively \cite{tan_skyrmion_2020, chen_unveiling_2022}. Magnetization data were acquired using an EZ11 vibrating sample magnetometer (VSM) from MicroSense™ in out-of-plane (OP) geometry. Magnetic imaging used an FEI Titan 80--300 TEM operated in Fresnel mode at 300~kV, with a defocus of $-2.4$~mm (see SM3),\cite{fultz_transmission_2012} and sample tilted 15° from normal incidence. The sample temperature was varied using a Gatan 636 cryogenic holder in automatic or bake-out mode, and OP fields were applied using the objective lens. Both experiments were performed over temperatures of 100--350~K, and applied OP magnetic fields ($\mu_0 H$) are herein referenced relative to the OP saturation field ($\mu_0 H_{\rm s}$).

\section*{FORC Magnetometry\label{sec:FORC}}
\begin{figure}
\centering\includegraphics[width=\columnwidth]{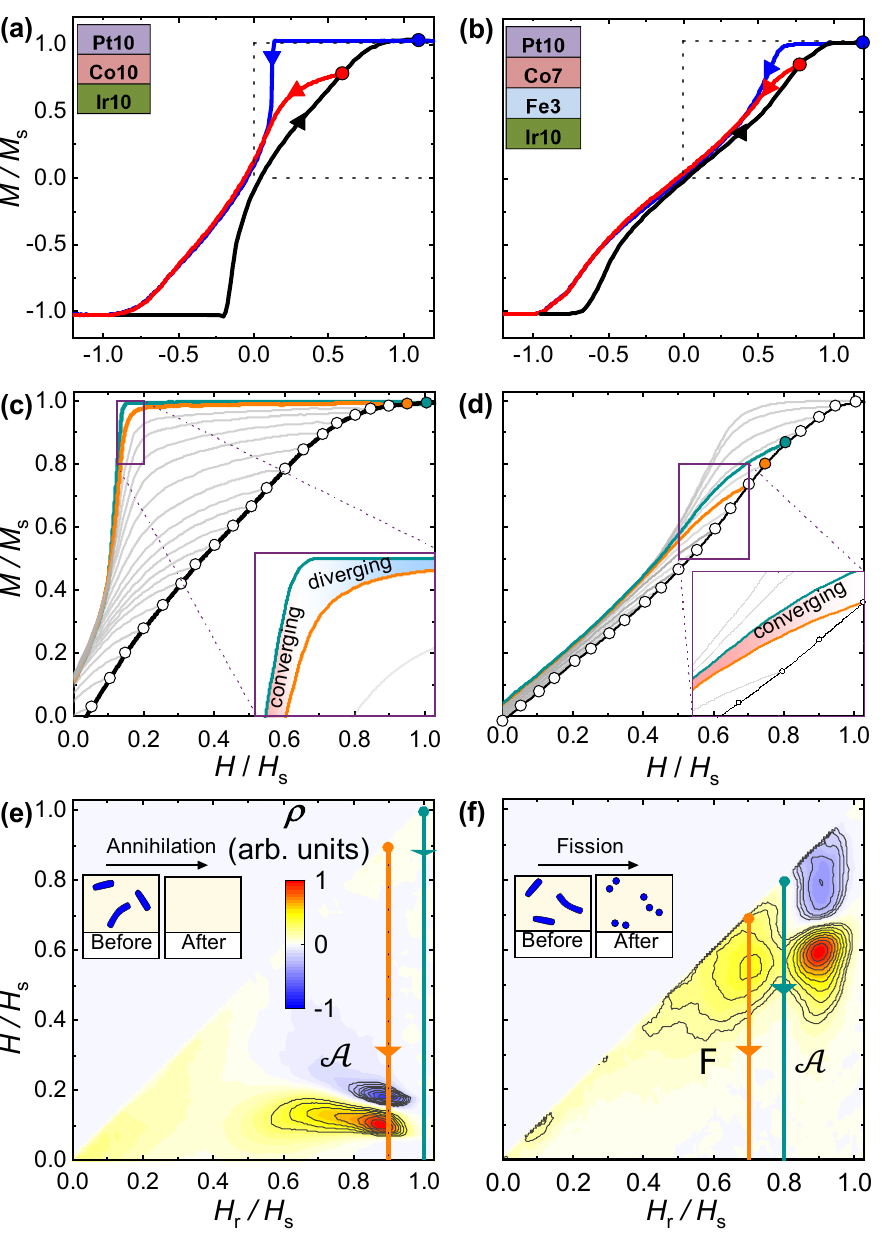}
\caption{\textbf{FORC magnetometry and irreversibility.}
\textbf{(a, b)} Out-of-plane (OP) magnetization FORCs for samples Fe(0)/Co(10) (a) and Fe(3)/Co(7) (b) at 300 K. Field was swept from $-H_{\rm s}$ to the reversal field, $H_{\rm r}$ (circles), and back to $-H_{\rm s}$. Arrows show sweep protocol for major (blue, $H_{\rm r}>H_{\rm s}$) and minor (red, $H_{\rm r}<H_{\rm s}$) loops. Dotted box is the region of interest for FORC analysis. 
\textbf{(c, d)} Set of FORCs at 300~K for both samples for selected $H_{\rm r}$ (circles, major loop in black). Zoomed insets show the $\sim 0.9 H_{\rm s}$ convergence-divergence (c) and $\sim 0.7 H_{\rm s}$ divergence (d) of proximate FORCs (green, orange).
\textbf{(e-f)} Color plot of FORC irreversibility, $\rho(H,H_{\rm r})$ (\ref{eqn:rho_definition}) for both samples. Irreversible processes, e.g. domain annihilation ($\mathcal{A}$, (e) inset) and fission ($\mathcal{F}$, (f) inset) present as non-zero $\rho$ features.} \label{fig1}\end{figure}

A FORC is a field segment of a minor hysteresis loop of magnetization, $M(H, H_{\rm r})$, characterized by a reversal field $H_{\rm r}$ ($\lesssim H_{\rm s}$) \cite{pike_characterizing_1999, davies_magnetization_2004, pike_first-order_2003}. The FORC field range for our work (up to $\pm 350$~mT) was determined by acquiring full OP hysteresis loops ($M(H)$) at each temperature. Subsequently, the samples were negatively saturated ($\mu_0 H < -1.2~\mu_0 H_{\rm s}$), and the field swept up to $\mu_0 H_{\rm r}$, and reversed, and data were acquired till $-10$~mT. A set of such FORCs were acquired at 2~mT intervals in $\mu_{0} H_{\rm r}$ over the stated temperatures for all samples. The resulting $M(H, H_{\rm r})$ data were fit to a second-order polynomial surface (see SM2), and were used to determine the irreversibility, $\rho(H, H_{\rm r})$ \cite{pike_characterizing_1999} 
\begin{equation}\label{eqn:rho_definition}
\rho(H,H_{\rm r}) = -\frac{1}{2}\frac{\partial^{2}M(H, H_{\rm r})}{\partial H\partial H_{\rm r}}.
\end{equation}
\begin{figure}
\centering\includegraphics[width=\columnwidth]{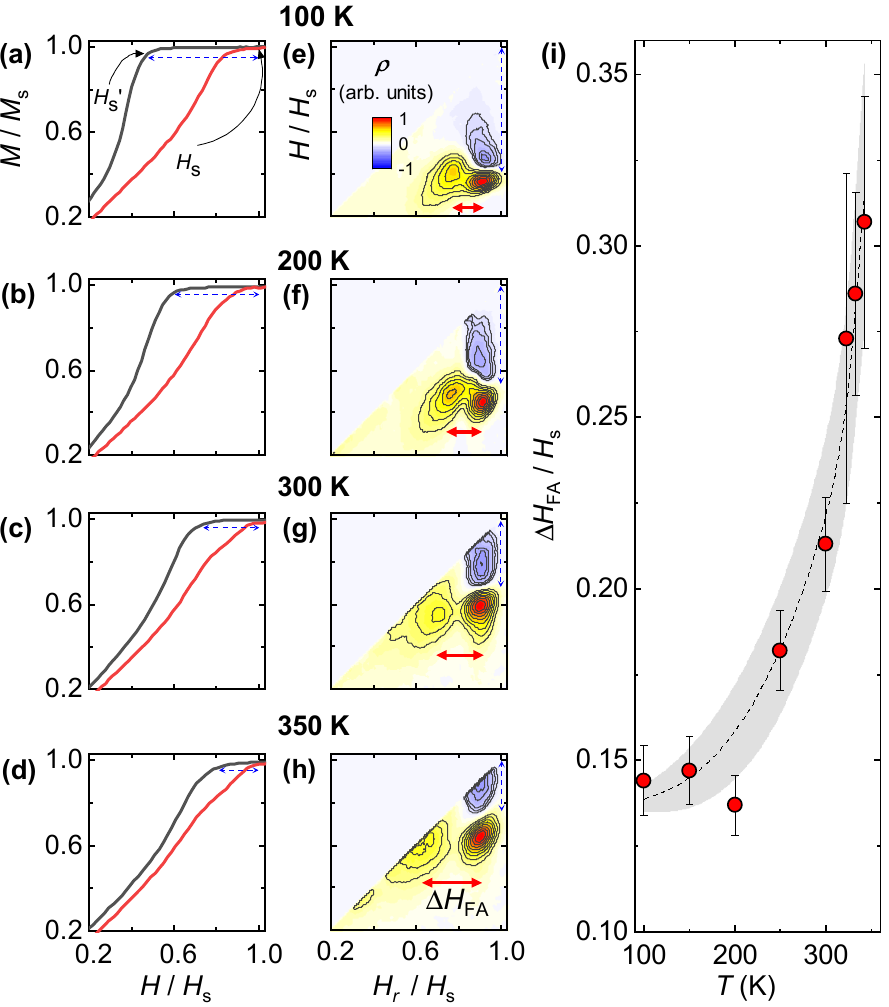}
\caption{\textbf{Thermal Evolution of Irreversibility.} 
\textbf{(a-d)} Zoom-ins of Fe(3)/Co(7) $M(H)$ loops of at 100, 200, 300, and 350~K respectively. $H_{\rm s}$ (red, polarized) and $H_{\rm s}'$ (black, unpolarized) indicate saturation fields, and blue arrow indicates their difference ($H_{\rm s}' – H_{\rm s}$).
\textbf{(e-h)} Color plots of FORC irreversibility, $\rho(H,H_{\rm r})$ at the respective temperatures. Red arrow indicates $\Delta H_{FA}$, the separation between $\mathcal{A}$ and $\mathcal{F}$ features (defined in (h)), dotted blue arrow indicates ($H_{\rm s}' – H_{\rm s}$, as in (a)).  
\textbf{(i)} Evolution of $\Delta H_{FA}$, normalised to $H_{\rm s}$ with temperature $T$. Error bars represents the total peak fit uncertainty (see SM2), the line is a guide-to-the-eye.}
\label{fig2}\end{figure}

Some magnetization reversal processes are smooth, e.g., spin canting or domain shrinking/expansion, and are can be reversed with applied field. Others, e.g. domain switching, are abrupt, and are therefore irreversible. FORC magnetometry is used to precisely identify and characterize such irreversible processes, notably via  $\rho (H, H_{\rm r})$ --- which is nonzero only in corresponding field regions \cite{pike_characterizing_1999, davies_magnetization_2004, pike_first-order_2003}. FORC had been used extensively study domain phenomenology in Co/Pt multilayers \cite{davies_magnetization_2004}, as illustrated in \ref{fig1}(a,c,e) for sample Fe(0)/Co(10) at RT. Here, the only discernible $\rho(H,H_{\rm r})$ feature is a peak-valley pair at $H_{\rm r} \sim 0.9~H_{\rm s}$. Labelled as $\mathcal A$, it arises from the irreversible annihilation of domains near saturation (\ref{fig1}(e), inset), due to sequential divergence and convergence of proximate FORCs (\ref{fig1}(c), inset), whose microscopic origin is well-established \cite{davies_magnetization_2004}. Recently, skyrmion formation in some samples was found to produce an additional $\rho(H,H_{\rm r})$ feature \cite{tan_skyrmion_2020}, shown in \ref{fig1}(f) for sample Fe(3)/Co(7). For $\kappa \gtrsim 1$ multilayers, individual stripes fission into multiple skyrmions at $H \sim 0.6~H_{\rm s}$ (\ref{fig1}(f), inset). This irreversible stripe-to-skyrmion fission imprints on $\rho(H,H_{\rm r})$ an additional ``$\mathcal F$'' peak, and ensues in dense skyrmion configurations and enhanced ZF stability \cite{lemesh_current-induced_2018, tan_skyrmion_2020, ang_electrical_2020}. In the absence of fission, e.g. for $\kappa < 1$ multilayers, stripes would instead shrink into individual skyrmions \cite{chen_unveiling_2022}.

\begin{figure*}[htb]
\includegraphics[width=\linewidth]{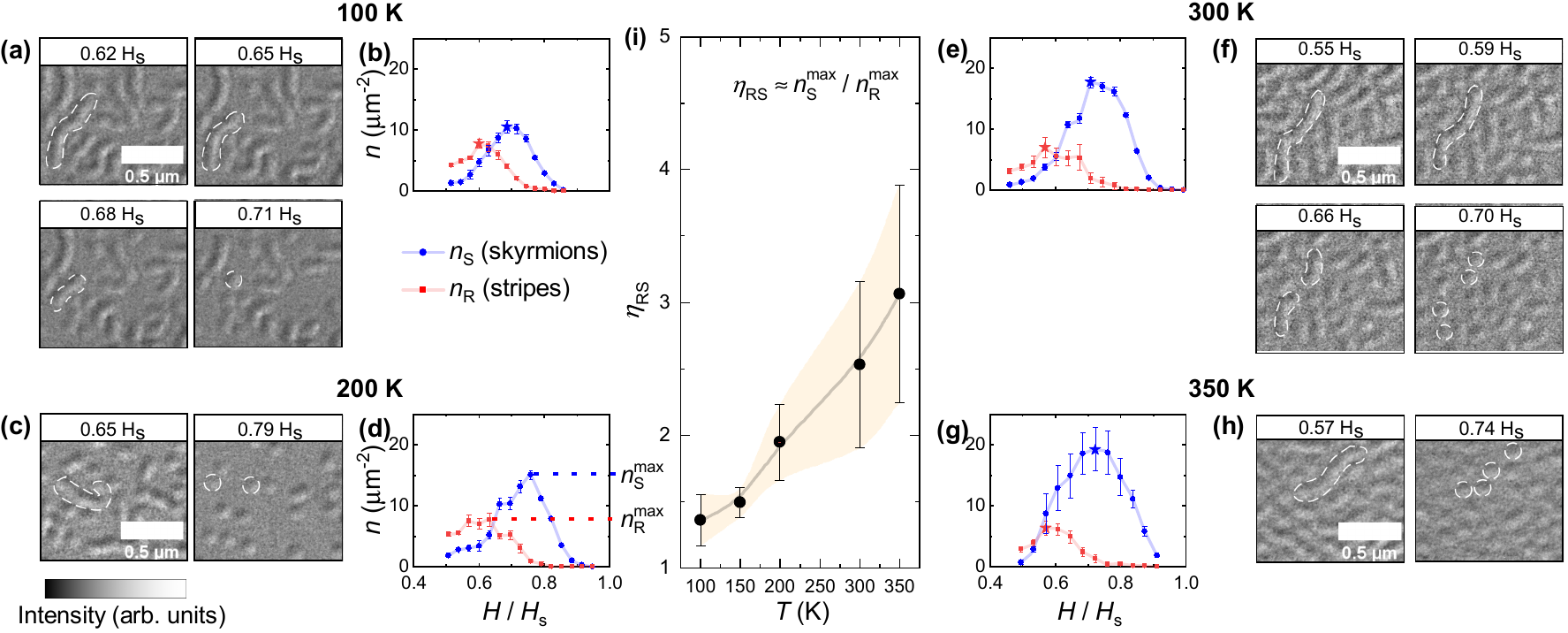}
\caption{\textbf{LTEM Imaged Texture Density Evolution.} 
\textbf{(a,c,f,h)} LTEM images of textural field evolution of Fe(3)/Co(7) at 100~K (a), 200 K (c), 300 K (f), and 350~K (h). Dotted white lines highlight individual stripe-to-skyrmion transitions. 
\textbf{(b,d,e,g)} Field dependence of measured skyrmion ($n_{\rm S}$, black) and stripe densities ($n_{\rm R}$, red). Stars represent maximum densities ($n_{\rm S}^{\rm max}$ and $n_{\rm R}^{\rm max}$, see (d)). Error bars represent confidence in texture identification (see SM3). 
\textbf{(i)} Average number of skyrmions formed per stripe, $\eta_{\rm RS}  \equiv n_{\rm S}^{\rm max}$ / $n_{\rm R}^{\rm max}$, plotted against temperature. Lines in (b,d,e,g,i) are guides-to-the-eye.} 
\label{fig3}\end{figure*}

We now turn to our key finding --- the temperature dependence of the stripe-to-skyrmion transition --- by examining its FORC signatures on Fe(3)/Co(7). Firstly, the $M(H)$ hysteresis loops vary discernibly over 100 -- 350~K (\ref{fig2}(a-d)). The extent of shear --- i.e., difference in saturation fields for polarized ($H_{\rm s}'$) and unpolarized ($H_{\rm s}$) curves (\ref{fig2}(a-d), blue arrows) --- is considerable at 100~K, and reduces with increasing temperature. This corresponds to the $H$-position of the annihilation ($\mathcal{A}$) peak in $\rho(H, H_{\rm r})$ plots (\ref{fig2}(e-g), blue arrows). At 100~K, $\mathcal{A}$ is distant from the $H = H_{\rm r}$ diagonal, and gradually inches towards it with increasing temperature.

More interestingly, the fission ($\mathcal{F}$) peak evolves markedly with temperature. At 100~K (\ref{fig2}(e)), $\mathcal{F}$ is nearly merged with $\mathcal{A}$. With increasing temperature (\ref{fig2}(f-h)), $\mathcal{F}$ breaks away from $\mathcal{A}$ and migrates towards the $H = H_{\rm r}$ diagonal. The separation of $\mathcal{F}$ and $\mathcal{A}$ peaks along $H_{\rm r}$ can be quantified by $\Delta H_{\mathcal{F}\mathcal{A}}$ (\ref{fig2}(e), inset, also see SM2) \cite{tan_skyrmion_2020}, and is indicative of the prevalence of fission. Indeed, $\Delta H_{\mathcal{F}\mathcal{A}}/H_{\rm s}$ increases sharply over 200-350~K (\ref{fig2}(i)), implying that at higher temperatures, fission occurs earlier relative to annihilation, and skyrmions thus formed are stable over a larger field range. Overall, the FORC results indicate increased dominance of the fission process with temperature.  

\section*{Lorentz Microscopy}

To directly visualize the stripe-to-skyrmion transition, we used LTEM, which images magnetic textures via Lorentz force deflection of incident electrons \cite{benitez_magnetic_2015, phatak_recent_2016}. Samples were negatively saturated ($-H_{\rm s}$), and as the field was swept to $+H_{\rm s}$, LTEM images were recorded over 8-10~$\mu$m fields-of-view. Subsequently, non-magnetic contributions were removed by subtracting a background image acquired above saturation  (see SM3). Expectedly, we observe a labyrinthine configuration at ZF, which transforms, with increasing field, into stripes, and then into 50-80~nm sized Néel skyrmions (see SM3) \cite{chen_unveiling_2022}. 
First, we examine the microscopic field evolution of an individual Fe(3)/Co(7) stripe (\ref{fig3}(a,c,e,g)). At 100~K (\ref{fig3}(a)), the highlighted stripe at 0.62~$H_{\rm s}$ shrinks in length as $H$ is increased, eventually into a single skyrmion at ~0.71~$H_{\rm s}$. In contrast, at 300~K (\ref{fig3}(f)), the highlighted stripe at 0.55~$H_{\rm s}$ fissions to produce four skyrmions at 0.7~$H_{\rm s}$.

The thermal evolution for individual stripes is consistent with statistical analyses of textures over $\sim$ 5 -- 7~$\mu$m fields-of-view (see SM3). \ref{fig3}(b,d,e,g) show the field evolution of the densities of stripes ($n_{\rm R}$) and skyrmions ($n_{\rm S}$) for Fe(3)/Co(7) over 100--350~K. We can quantify the efficacy of the fission process by $\eta_{\rm RS}(T)$, the average number of skyrmions formed per stripe. Here, we can approximate $\eta_{\rm RS}$ $\approx n_{\rm S}^{\rm max}$ / $n_{\rm R}^{\rm max}$, where  ($n_{\rm S}^{\rm max}$ and $n_{\rm R}^{\rm max}$ are the maximum skyrmion and stripe densities at a given temperature (see \ref{fig3}(d)). As shown in \ref{fig3}(i), $\eta_{\rm RS}$ displays a monotonic rise with temperature, from  $\sim 1$ at 100~K to $\sim$ 3 at 350~K. This suggests that at higher temperatures each stripe fissions into more skyrmions. Meanwhile, the suppression of fission at lower temperatures results in the shrinking of stripes to skyrmions. Finally, we note that the increase in $\eta_{\rm RS}$ expectedly lead to higher $n_{\rm S}^{\rm max}$, i.e. dense skyrmion configurations \cite{tan_skyrmion_2020}.   

\label{sec:FORC_LTEM}
\begin{figure}
\includegraphics[width = \columnwidth]{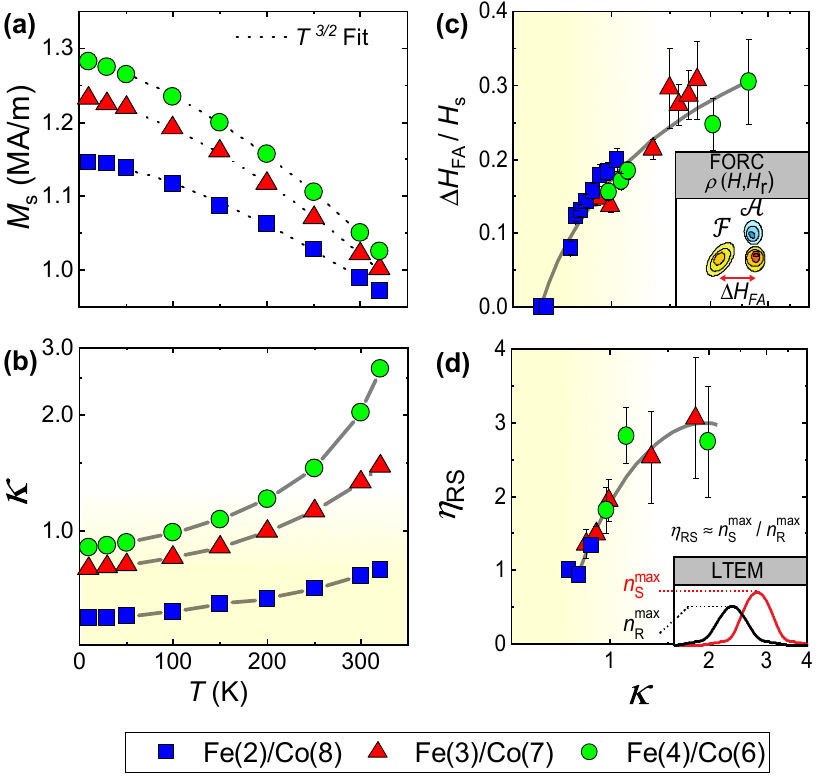}
 \caption{\textbf{Thermodynamic evolution of stripe-skyrmion transition.} 
\textbf{(a,b)} Temperature dependence of saturation magnetisation $M_{\rm s}$ (a) and estimated $\kappa$ (b, see text) for the three studied samples. For (a), dashed lines indicate vertical offsets (0.05~MA/m each), dotted curves represent Bloch $T^{3/2}$ fits. 
\textbf{(c,d)} Compiled variation of $\Delta H_{\mathcal{F}\mathcal{A}}$ (from FORC $\rho (H,H_{\rm r})$) and $\eta_{\rm RS}$ ($\equiv n_{\rm S}^{\rm max}/n_{\rm R}^{\rm max}$, from LTEM) with $\kappa_{\rm est}$ across samples and temperatures. Schematic insets (c-d) show definitions of $\Delta H_{\mathcal{F}\mathcal{A}}$ (for (c)) and $n_{\rm S}^{\rm max}$, $n_{\rm R}^{\rm max}$ (for (d)) respectively. Lines in (b-d) are guides-to-the-eye.}
\label{fig4}\end{figure}

Together, these results on Fe(3)/Co(7) indicate that as temperature increases, stripe fission becomes the dominant skyrmion formation mechanism. Temperature variations are known to affect textural transitions in chiral multilayers via (a) thermal fluctuations \cite{lemesh_current-induced_2018}, and/or (b) magnetic parameter evolution \cite{je_targeted_2021}. To delineate these effects, we perform similar experiments over 100 -- 350~K on two more samples --- Fe(2)/Co(8) and Fe(4)/Co(6) --- providing a wide range of magnetic parameters \cite{tan_skyrmion_2020,chen_unveiling_2022}. Consistently across samples, the saturation magnetization $M_{\rm s}$ follows the $T^{3/2}$ Bloch's law (\ref{fig4}(a)), and is expected to to influence the key magnetic interactions as follows: $A(T) \propto M_{\rm s}(T)^{1.8}$; $D(T) \propto M_{\rm s}(T)^{1.8}$, and uniaxial anisotropy $K_{\rm u}(T) \propto M_{\rm s}(T)^{2.6}$ (see SM1) \cite{callen_present_1966, tomasello_origin_2018, moreau-luchaire_additive_2016, nembach_linear_2015, je_targeted_2021}. Using these relations, we calculate the expected thermal evolution of $\kappa$ ($\simeq \kappa_{\rm est}$) across samples. \ref{fig4}(b) shows that $\kappa_{\rm est}$ increases monotonically with temperature, with marked variation on either side of unity ($\sim 0.6-2.8$).  We note that for all three samples, across the entire temperature range of investigation, $D$ is approximately 3 -- 6 times greater than the DW anisotropy energy, \cite{kim_correlation_2018, franke_switching_2021} which consistently results in the stabilization Néel DWs despite the lower magnetostatic energy of Bloch DWs.

\ref{fig4}(c-d) show the compiled evolution of FORC-measured $\Delta H_{\mathcal{F}\mathcal{A}}/H_{\rm s}$ and LTEM-measured stripe-skyrmion ratio, $\eta_{\rm RS}$ with $\kappa_{\rm est}$. Consistently across samples and temperatures, we find that as $\kappa_{\rm est}$ increases from $\sim 0.6$ to $\sim 2.8$, both $\Delta H_{\mathcal{F}\mathcal{A}}/H_{\rm s}$ ($\sim 0$ to $\sim 0.3$) and $\eta_{\rm RS}$ ($\sim 1$ to $\sim 3$) show increasing trends. 
While $\Delta H_{\mathcal{F}\mathcal{A}}/H_{\rm s} (T)$ and $\eta_{\rm RS} (T)$ also grow monotonically for each sample (see SM4), \ref{fig4}(c-d) show that the $\kappa$-dependence of both characteristics collapses onto a single curve across samples and temperatures. It is likely that magnetic parameters --- not thermal excitations \cite{lemesh_current-induced_2018} --- drive the temperature evolution of the stripe-to-skyrmion transition, and $\kappa$ plays an important role in the fission process. 

\section*{Atomistic Calculations}
\begin{figure}
\includegraphics[width = \columnwidth]{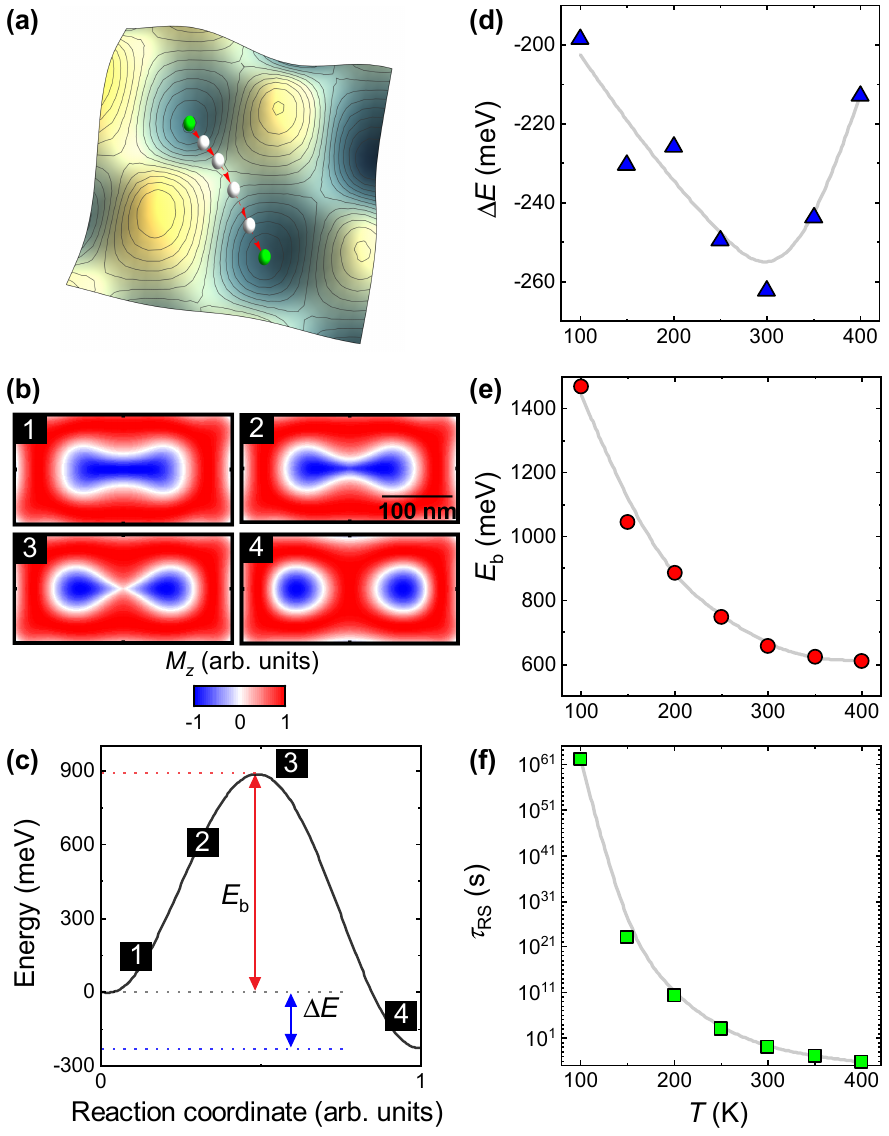}
\caption{\textbf{Stripe-Skyrmion Transition Energetics.} 
\textbf{(a)} Schematic energy surface, showing the minimum energy path (red arrows) between two quasi-equilibrium states (green dots). White dots show intermediate states. 
\textbf{(b)} GNEB simulated magnetization images of stripe-to-skyrmion fission for Fe(3)/Co(7) parameters at $T = 200$~K, showing initial (1), intermediate (2-3), and final (4) magnetization states.
\textbf{(c)} Energy profile for the transition in (b), with positions of states (1-4) overlaid. The energy barrier, $E_{\rm b}$ (red arrow) and net energy difference, $\Delta E$ (blue arrow) – are indicated. 
\textbf{(d-f)} Plots of $\Delta E(T)$ (d), $E_{\rm b}(T)$ (e), and transition lifetime ($\tau_{\rm RS} (T)$ ((f), from \ref{eqn:lifetime}) against temperature for stripe-to-skyrmion fission for Fe(3)/Co(7) parameters. Solid lines in (d-f) are guides-to-the-eye.}
\label{fig5}\end{figure}

The fission of a stripe into multiple skyrmions must overcome topological protection and the associated energy barrier \cite{heil_universality_2019}. To understand the fission energetics, we turn to GNEB calculations performed using the Fidimag package \cite{bisotti_fidimag_2018}. GNEB calculates the minimum energy path for transitions between two fixed quasi-equilibrium magnetic states (schematic: \ref{fig5}(a)), while implicitly constraining magnetic moments to fixed magnitude \cite{jnsson_nudged_1998,henkelman_climbing_2000,bessarab_method_2015}. The energy, $E$ of the magnetic configurations is calculated using the atomistic Heisenberg Hamiltonian and referenced to the initial state \cite{cortes-ortuno_thermal_2017}.

Here the initial state was chosen to be a stripe (\ref{fig5}(b): (1)) and the final state to be two-skyrmions (\ref{fig5}(b): (4)), and these states were first relaxed using the atomistic Landau-Lifshitz-Gilbert equation under a 50~mT magnetic field \cite{bessarab_method_2015, bisotti_fidimag_2018}. Subsequently GNEB was used to determine the transition energy profile (shown in \ref{fig5}(c)). Temperature dependence was incorporated using rescaled magnetic interactions  (scaling laws in \ref{fig4}). First, \ref{fig5}(d) shows that the transition energy difference, $\Delta E < 0$ at all temperatures, i.e. the two-skyrmion state is consistently energetically favoured to the stripe state. While $\Delta E (T)$ does show up to ~20$\%$ variation, the trend is non-monotonic. In contrast, the transition energy barrier $E_{\rm b}(T)$ (\ref{fig5}(e)) exhibits a sharp ($\sim 3\times$), monotonic reduction with increasing temperature. Thus, we posit the direct association of $E_{\rm b}(T)$ with the observed increase in stripe-to-skyrmion fission activity.

We estimate the stripe-to-skyrmion transition lifetime, $\tau_{\rm RS}$, using Néel-Arrhenius relaxation theory as \cite{cortes-ortuno_thermal_2017}
\begin{equation}\label{eqn:lifetime}
\tau_{\rm RS} =\tau_{0} \exp\left(E_{\rm b}/k_{\rm B}T\right), 
\end{equation}
where $\tau_0$ is the attempt period (typically $\sim 10^{-9} -- 10^{-12}$~s \cite{cortes-ortuno_thermal_2017,schrefl_micromagnetic_2001, aharoni_introduction_2001}) and $k_{\rm B}$ is the Boltzmann constant. With $\tau_0$ $\sim$ 10$^{-12}$~s, we find that $\tau_{\rm RS}$ reduces dramatically from $\gg 10^{50}$~s at 100~K to $\sim 0.1$~s at 300~K. This order-of-magnitude transition rate estimate supports the hypothesis that over 100--350~K, stripe-to-skyrmion fission for Fe(3)/Co(7) evolves from kinematically forbidden to rapidly accessible. In contrast, if we assume a temperature-independent energy barrier fixed at its 100~K value, the transition lifetime remains extremely long at $\tau_{RS}>10^{10}~s$ at 300~K, which contradicts our experimental observations. This confirms that thermal fluctuations alone cannot account for the observed $T$-dependence of fission and that $\kappa(T)$-driven energy barrier reduction is crucial to enabling fission kinematics, which, in turn, manifest in the temperature dependence of $\Delta H_{\mathcal{F}\mathcal{A}}$ and $\eta_{\rm RS}$.  

\section*{Summary}

In summary, we have presented detailed thermodynamic and microscopic evidence on the temperature evolution of stripe-to-skyrmion fission in chiral multilayers. The increased prominence of the magnetometry fission peak  and enhanced fission efficacy in LTEM are both well-described by the thermal evolution of the DW stability parameter $\kappa$ across samples. Atomistic calculations establish that this arises from the sharp reduction of the energy barrier to stripe fission at elevated temperatures. Our work provides timely insights on the thermodynamics of the stripe-to-skyrmion transition. On one hand, while anisotropy and temperature are known to independently enhance skyrmion density \cite{soumyanarayanan_tunable_2017, desautels_realization_2019, tan_skyrmion_2020, raju_evolution_2019}, these results establish a much-needed bridge between the thermodynamics and microscopics. On the other hand, these thermodynamic effects can be exploited, e.g. via controlled temperature cycles, to generate skyrmion textures \cite{lemesh_current-induced_2018, brock_current-induced_2020, wang_electric-field-driven_2020, ang_electrical_2020, je_targeted_2021}. Our framework may be particularly relevant to manipulating chiral spin texture ensembles for unconventional computing \cite{zazvorka_thermal_2019}.

\noindent We acknowledge the support of the National Supercomputing Centre (NSCC), Singapore, for computational resources. This work was supported by the SpOT-LITE program (Grant No. A18A6b0057), funded by Singapore's RIE2020 initiatives. E.C. was partially supported by the NUS Resilience and Growth Traineeship Programme. 

\phantomsection\addcontentsline{toc}{section}{\refname}
\bibliography{SKFORC}


\normalsize
\clearpage
\onecolumngrid
\begin{center}
\textbf{\Large Supplementary Materials}
\end{center}
\setcounter{section}{0}
\setcounter{figure}{0}
\setcounter{table}{0}
\setcounter{equation}{0}

\setcounter{secnumdepth}{4}
\renewcommand\thesection{S\arabic{section}}
\renewcommand{\theparagraph}{S\arabic{section}\alph{paragraph}}
\makeatletter\@addtoreset{paragraph}{section}\makeatother
\makeatletter\def\p@paragraph{}\makeatother
\renewcommand{\thefigure}{S\arabic{figure}}
\renewcommand{\theequation}{S\arabic{equation}}
\renewcommand{\thetable}{S\arabic{table}}


\renewcommand{\ref}[1]{\autoref{#1}}
\renewcommand{\figureautorefname}{Figure}
\renewcommand{\equationautorefname}{Equation}
\renewcommand{\tableautorefname}{Table}
\renewcommand{\sectionautorefname}{\S}
\renewcommand{\subsectionautorefname}{\S}
\renewcommand{\paragraphautorefname}{\S}

\setcounter{tocdepth}{1}
\makeatletter\def\l@section{\@dottedtocline{1}{0.6em}{1.5em}}\makeatother
\makeatletter\def\l@paragraph{\@dottedtocline{4}{1.5em}{1.8em}}\makeatother
\makeatletter\def\l@figure{\@dottedtocline{1}{0.6em}{1.8em}}\makeatother

\linespread{1.25}
\def\arraystretch{1.5}
\setlength{\parskip}{1.5ex plus0.2ex minus0.2ex} 
\setlength{\abovecaptionskip}{4pt}\setlength{\belowcaptionskip}{-4pt}
\setlength{\abovedisplayskip}{0ex}\setlength{\belowdisplayskip}{0ex}
\setlength{\abovedisplayshortskip}{0ex}\setlength{\belowdisplayshortskip}{0ex}

\titleformat{\section}{\large\bfseries\scshape\filcenter}{\thesection.}{1em}{#1}[{\titlerule[0.5pt]}]
\titlespacing*{\section}{0pt}{1ex}{1ex}
\titleformat{\subsection}{\bfseries\sffamily}{\thessubsection.}{0em}{#1}
\titlespacing{\subsection}{0pt}{0.5ex}{0.5ex}
\titleformat{\paragraph}[runin]{\sffamily\bfseries}{}{-1.2em}{#1.}
\titlespacing*{\paragraph}{1.25em}{2ex}{0.4em}[] 

\section{Magnetic Properties and Temperature Dependence}

The Ta(40)/Pt(50)/[Ir(10)/Fe($x$)/Co($y$)/Pt(10)]$_{14}$/Pt(20) multilayer samples used in this work (layer thickness in angstroms in parentheses) and their room temperature (RT) magnetic properties are tabulated in \ref{tab:magnetic_props}. The RT magnetic properties of multilayer samples with identical compositions sputtered using similar deposition parameters have been extensively characterized in our previous works \cite{tan_skyrmion_2020, chen_unveiling_2022}. Here, the RT saturation magnetisation, $M_{\rm s}$, and effective anisotropy,  $K_{\rm eff}$, were determined from VSM measurements. Meanwhile, the RT values of exchange stiffness ($A_{\rm est}$) and interfacial Dzyaloshinskii-Moriya interaction iDMI ($D_{\rm est}$) were derived from our previous works, which used identical stack compositions  \cite{tan_skyrmion_2020, chen_unveiling_2022}. 

\begin{table}[h]
\begin{tabular}{|l|l|c|c|c|c|c|}
\hline
\multirow{2}{*}{\textbf{Acronym}} & \multirow{2}{*}{\textbf{Stack Composition}} &
$\boldsymbol{M_{\rm s}}$ &  $\boldsymbol{K_{\rm eff}}$ & 
$\boldsymbol{D_{\rm est}}$ & $\boldsymbol{A_{\rm est}}$ & \multirow{2}{*}{\textbf{$\boldsymbol{\kappa}$}} \\
&  &  (MA/m) & (MJ/m$^3$) & (mJ/m$^2$) & (pJ/m) & \\
\colrule\hline
Fe(0)/Co(10) & [Ir(10)/Co(10)/Pt(10)]$_{14}$ & 1.16 & 0.60 & 1.2 & 17.8 & 0.3 \\
\hline
\textbf{Fe(2)/Co(8)} &[Ir(10)/Fe(2)/Co(8)/Pt(10)]$_{14}$ & 1.14 & 0.26 & 0.9 & 12.8 & 0.4 \\
\hline
\textbf{Fe(3)/Co(7)} & Ir(10)/Fe(3)/Co(7)/Pt(10)]$_{14}$ & 1.11 & 0.095 & 1.7 & 13.2 & 1.2 \\
\hline
\textbf{Fe(4)/Co(6)} & [Ir(10)/Fe(4)/Co(6)/Pt(10)]$_{14}$ & 1.05 & 0.058 & 1.9 & 13.6 & 1.6 \\
\hline
\end{tabular}
\caption{\label{tab:magnetic_props}%
\textbf{Sample Compositions and RT Magnetic Properties.} 
List of samples used in this work, with Fe($x$)/Co($y$) acronyms, active stack compositions (layer thickness in angstroms in parentheses), and key RT magnetic properties --- $M_{\rm s}$, $K_{\rm eff}$, $D_{\rm est}$ , $A_{\rm est}$, and $\kappa$. Note: Fe(0)/Co(10) is used only for illustrative purposes (manuscript Fig. 1(a,c,e)).}
\end{table}

\begin{figure}[h]\centering
\includegraphics[width=16cm]{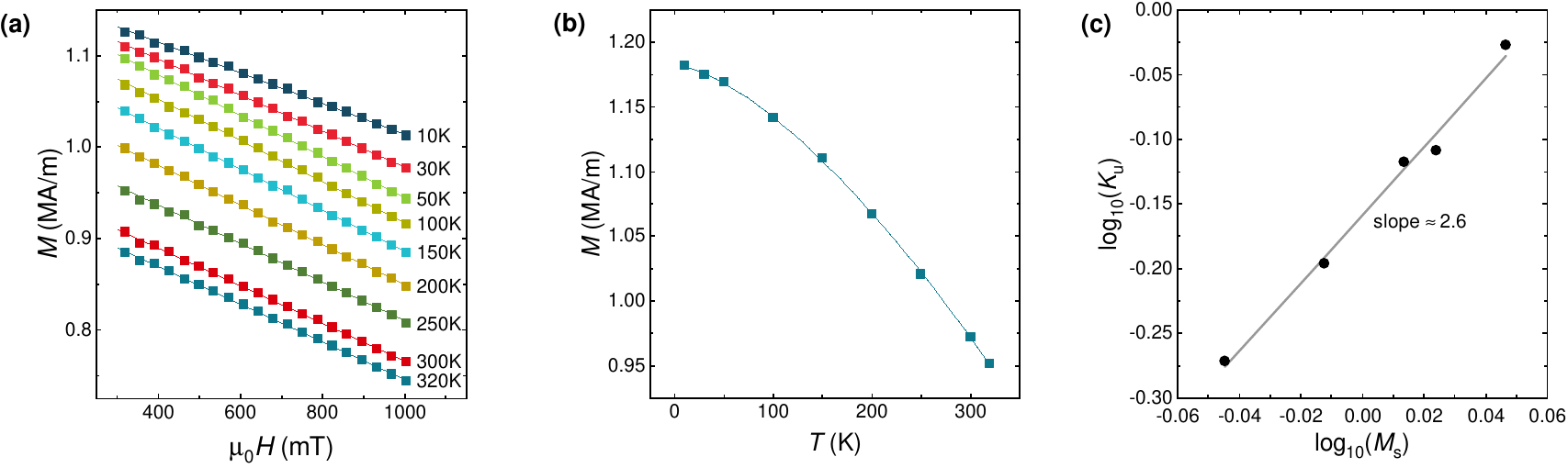}
\caption{\textbf{Measured $T$-dependence of Fe(3)/Co(7) magnetic parameters.} 
\textbf{(a)} Out-of-plane $M(H)$ measurements over $H>H_{\rm s}$, performed for 10~K to 320~K. Lines show linear fits, whose $y$-intercepts correspond to $M_{\rm s}$.  
\textbf{(b)} The $T$-dependence of $M_{\rm s}$, as determined from (a). Line shows a Bloch law ($T^{3/2}$) fit. 
\textbf{(c)} Logarithmic scatter plot of the measured uniaxial anisotropy, $K_{\rm u} = K_{\rm eff} + \mu_0 M_{\rm s}^2/2$ against $M_{\rm s}$ across temperatures. Line shows a linear fit with slope $\approx 2.6$.}
\label{fig:TDep_MagProps}
\end{figure}

To determine the temperature ($T$)-dependence of magnetic parameters ($A_{\rm est}$, $D_{\rm est}$) for manuscript Fig. 4, we first measured the $T$-dependence of $M_{\rm s}$ using an MPMS-SQUID magnetometer from Quantum Design\texttrademark. To correctly account for the $T$-dependent diamagnetic contribution of the substrate, we performed at each temperature a magnetic field sweep at fields $H$ greater than the saturation field, $H_{\rm s}$, of the sample . As shown in \ref{fig:TDep_MagProps}(a) for sample Fe(3)/Co(7), each field sweep thus acquired was fit to a straight line, and the $M_{\rm s}$ was read off as the $y$-intercept. Subsequently, as shown in \ref{fig:TDep_MagProps}(b), the $M_{\rm s}(T)$ data was fit to the Bloch $T^{3/2}$ law, 
$M_{\rm s}(T) = M_0(1 - \beta T^{3/2})$.
$M_{\rm s}(T)$ was then used to estimate the $T$-dependence of other magnetic parameters ($A_{\rm est}$, $D_{\rm est}$) using established scaling relations detailed in the manuscript.

Finally, the $T$-dependence of anisotropy was accounted for by performing in-plane and out-of-plane $M(H)$ measurements over the temperature range of interest. \ref{fig:TDep_MagProps}(c) shows a plot of the empirically measured scaling relationship of uniaxial anisotropy $K_{\rm u}$ with $M_{\rm s}$. The observed relationship --- $K_{\rm u} \propto M_{\rm s}^{2.6}$ --- is similar to previous works which estimated $K_{\rm u} \propto M_{s}^{3}$ \citep{callen_present_1966,moreau-luchaire_additive_2016,tomasello_origin_2018}.

\section{FORC Magnetometry Measurements and Analysis\label{sec:magnetometry_analysis}}
\begin{figure}[h]\centering
\includegraphics[width=16.5cm]{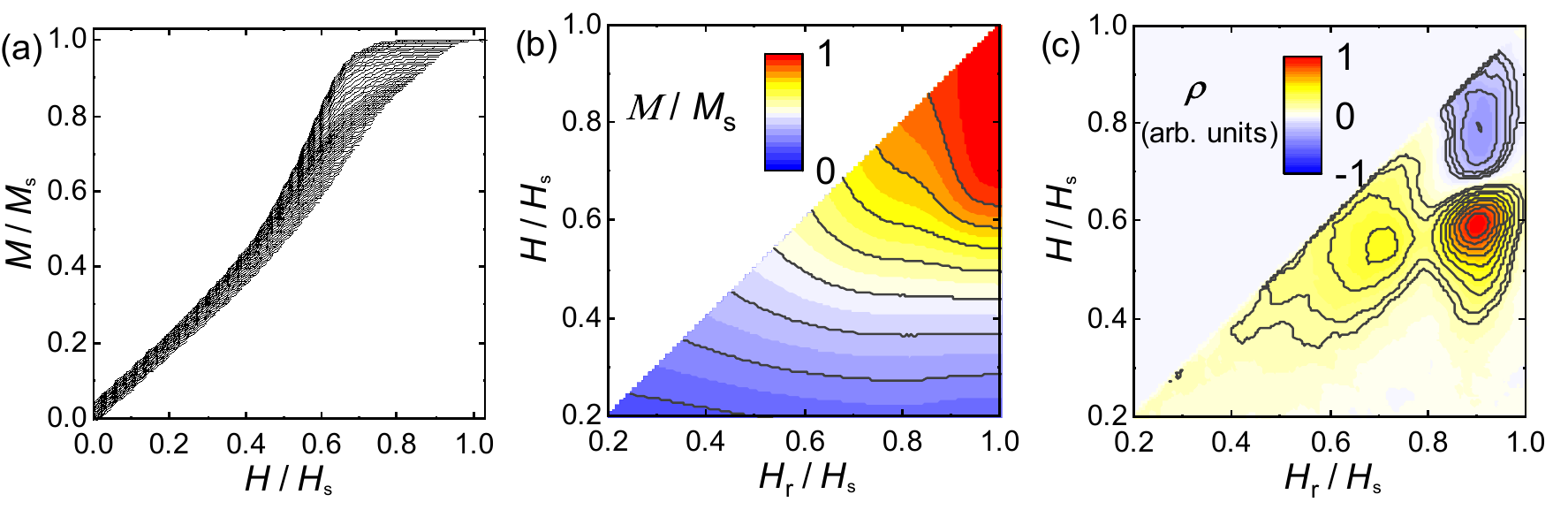}
\caption{\textbf{VSM and FORC measurements.} \textbf{(a).} Set of FORCs obtained for sample Fe(3)Co(7), with $H_r$ and H at intervals of 2~mT. \textbf{(b).} Color plot of magnetisation for the FORC data in (a), projected on H, $H_r$ axis. (\textbf{c).} Color plot of $\rho$, the FORC irreversibility, derived from the data in (a-b).}
\label{fig:FORC_method}\end{figure}

The color plots of FORC irreversibility, $\rho(H,H_{\rm r})$ --- shown in manuscript Fig. 1(e-f), and reproduced in \ref{fig:FORC_method}(c) --- were obtained by processing the set of VSM-measured FORCs, shown e.g. in \ref{fig:FORC_method}(a). First, the measured magnetization, $M(H,H_{\rm r})$ (\ref{fig:FORC_method}(b)), was fit to a second-order polynomial surface \citep{pike_characterizing_1999,pike_first-order_2003}:  
\begin{equation}\label{eqn:FORC_polyfit} 
M(H,H_{\rm r}) = a_1 + a_2 H_{\rm r} + a_3 H + a_4 H_{\rm r}^2 + a_5 H^2 + a_6 H_{\rm r} H.
\end{equation} 
The number of points used for fitting was determined by a smoothing factor, which is 5 in our work. Following from \ref{eqn:FORC_polyfit}, $\rho(H,H_{\rm r})$, defined as
\begin{equation}
\rho(H,H_{\rm r}) = -\frac{1}{2}\frac{\partial^{2}M(H, H_{\rm r})}{\partial H\partial H_{\rm r}},
\end{equation}
is given by $-a_6$.

To quantify the separation between the prominent irreversible features --- $\mathcal{A}$ and $\mathcal{F}$ --- in FORC $\rho(H,H_{\rm r})$ plots ($\Delta H_{\mathcal{F}\mathcal{A}}$, see manuscript Fig. 2), it is necessary to identify the centres of the respective features. This was accomplished using an algorithm written in Python code, as follows. First, the FORC distribution was binarized to identify $\rho > 0$ regions. Next, a watershed segmentation procedure was implemented for FORC distributions that visually exhibited overlap of two distinct features. The centres of these features in ($H, H_{\rm r}$) space were then identified by appropriately taking the weighted averages, thus enabling the determination of $\Delta H_{\mathcal{F}\mathcal{A}}$. Meanwhile, the standard deviation thus obtained indicates the spread of these features in ($H, H_{\rm r}$) space. 

\section{Lorentz TEM Measurements and Analysis\label{sec:LTEM}}
\begin{figure}[h]\centering
\includegraphics[width=15cm]{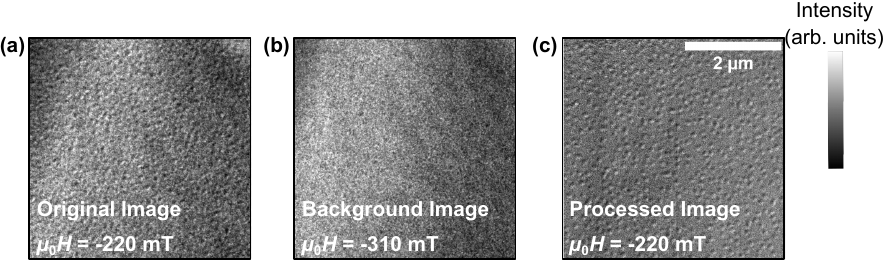}
\caption{\textbf{Background Removal for LTEM Images.} 
\textbf{(a-b)} As-acquired LTEM image at $\mu_0 H \simeq -220$~mT, depicting a sparse skyrmion configuration (a), and corresponding ``background'' (saturated) image at $\mu_0 H \simeq -310$~mT (b) for Fe(3)/Co(7) at RT. 
\textbf{(c)} Processed image for $\mu_0 H \simeq -220$~mT, obtained by subtracting (b) from (a) after translating and rotating (a) to align with (b), with visibly enhanced magnetic texture contrast c.f. (a). }
\label{fig:LTEM_BgSub}
\end{figure}

Phase shift $\gamma(k)$ for a scattered beam is caused by spherical aberration and defocus, where $\gamma(k)=2\pi(C_s \lambda k^4/4+\Delta f\lambda k^2/2)$ \cite{fultz_transmission_2012}. Dark rings are visible in the diffractogram when the phase shift is a multiple of $\pi$, i.e. $\gamma(k)=n\pi$. The defocus values reported in this work were determined from the intercept of a plot of $n/k^2$  against $k^2$ from a calibrated diffractogram taken at a particular defocus.

To enhance the magnetic contrast in Lorentz TEM images and improve the reliability of the statistical analysis of magnetic textures, we first perform a background subtraction procedure for each LTEM image to remove prominent sources of non-magnetic contrast. The procedure utilizes a reference, or ``background'' image -- without magnetic textures (i.e. captured at saturation) -- over approximately the same field-of-view. This allows the removal of contrast from non-magnetic sources, such as due to structural inhomogeneity of the film, waviness of the substrate etc. This process is accomplished, using a custom-written Python code, by first aligning the two images to correct for translational and rotational drifts. Next, simple pixel-wise subtraction is performed. The importance of the background subtraction process is illustrated in \ref{fig:LTEM_BgSub}. The contrast of magnetic textures is much more prominent in the processed image (\ref{fig:LTEM_BgSub}(c)) c.f. the original image (\ref{fig:LTEM_BgSub}(a)).

\begin{figure}[h]\centering
\includegraphics[width=7.5cm]{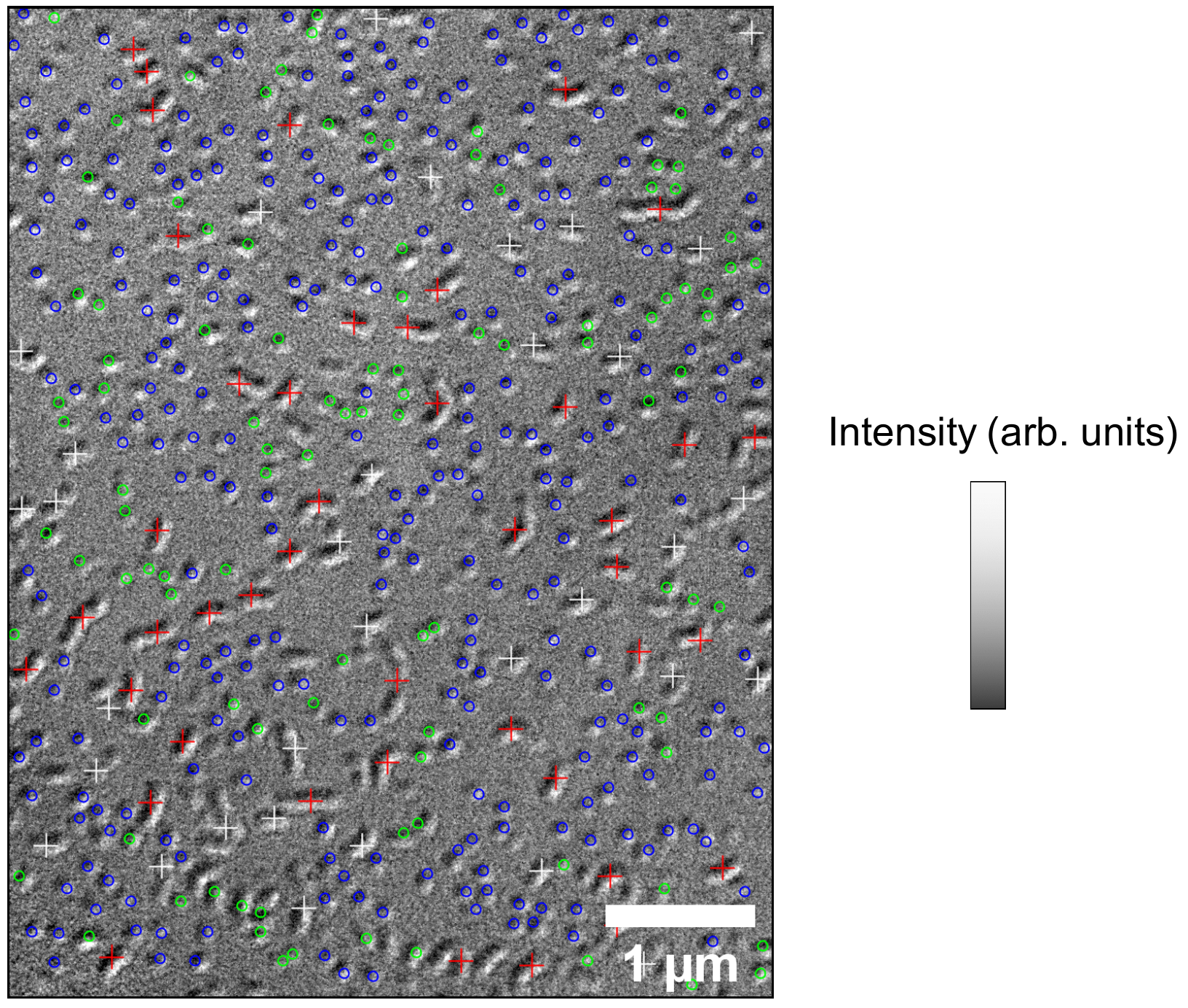}
\caption{\textbf{Identifying \& Counting Magnetic Textures.} Representative processed LTEM image of Fe(3)/Co(7) acquired at RT, $\mu_0 H \simeq -240$~mT, with a mixed configuration with stripes and skyrmions over a $\sim 30$~$\mu{\rm m}^2$ area. Textures identified as skyrmions and stripes are annotated with circles and crosses respectively. For skyrmions (stripes), blue and green (red and white) colors indicate certain and uncertain identifications of the respective magnetic textures.}
\label{fig:LTEM Counting}
\end{figure}

\begin{figure}[h]\centering
\includegraphics[width=17cm]{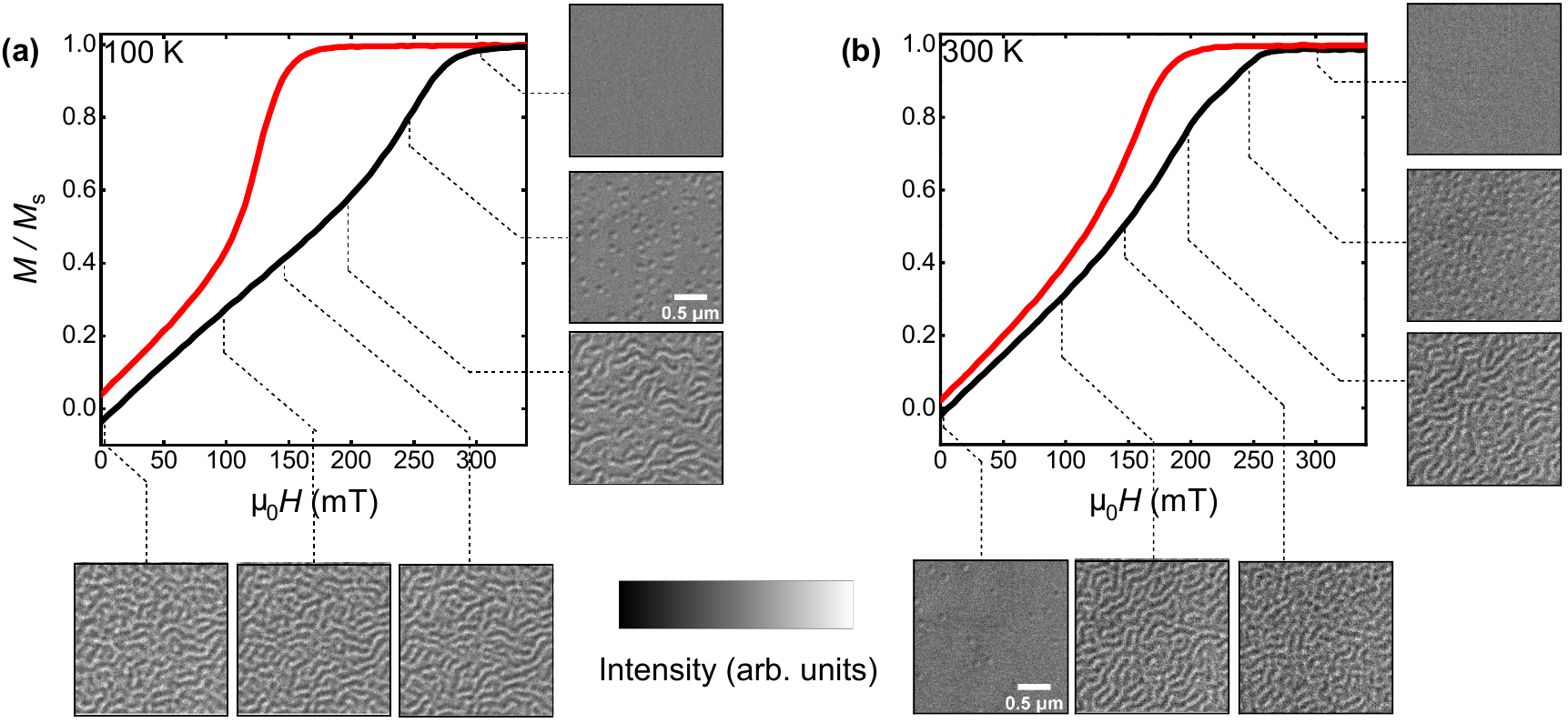}
\caption{\textbf{Field Evolution of LTEM-Imaged Textures} 
$M(H)$ curves of Fe(3)/Co(7), showing field-polarized (red) and unpolarized (black) sweeps measured by VSM at 100~K (a) and 300~K (b) respectively. Insets show series of LTEM images recorded along the unpolarized (black) curves at fields corresponding to dashed black lines.}
\label{fig:LTEM_FieldEvol}
\end{figure}

To quantify the field and temperature evolution of magnetic texture densities (see e.g. manuscript Fig. 3), every domain observed in LTEM images recorded over $H \sim (0.5 -- 1) \cdot H_{\rm s}$ (, \ref{fig:LTEM_FieldEvol}) was visually identified as a skyrmion or a stripe. An example of the outcome of such texture identification is shown in \ref{fig:LTEM Counting}. To reflect our confidence in the identification of skyrmions and stripes within dense domain backgrounds, the final skyrmion and stripe counts are obtained from weighted sums. Specifically, domains that are certain skyrmions or stripes (\ref{fig:LTEM Counting}: blue circles, red crosses) contribute their full value of unity to the respective domain counts. Meanwhile, domains of uncertain identity (\ref{fig:LTEM Counting}: green circles, white crosses) contribute half their value to their respective counts, while the other half of their magnitude is added to the corresponding error bars. 

The LTEM images shown in manuscript Fig. 3 focus on elucidating the microscopics of the stripe-to-skyrmion transition. In \ref{fig:LTEM_FieldEvol}, we present representative LTEM data  acquired on Fe(3)/Co(7) at 100~K and 300~K over a larger field range. In both cases, at zero field, we observe a labyrinthine stripe configuration, which transforms, with increasing field, first into stripes, and eventually into skyrmions. Notably, the density of magnetic textures, esp. skyrmions, is much higher for the 300~K data (\ref{fig:LTEM_FieldEvol}(b)) compared to that at 100~K (\ref{fig:LTEM_FieldEvol}(a)).

\section{Sample Dependence Analysis\label{sec:skyrmion stability}}

\begin{figure}[h]\centering
\includegraphics[width=12cm]{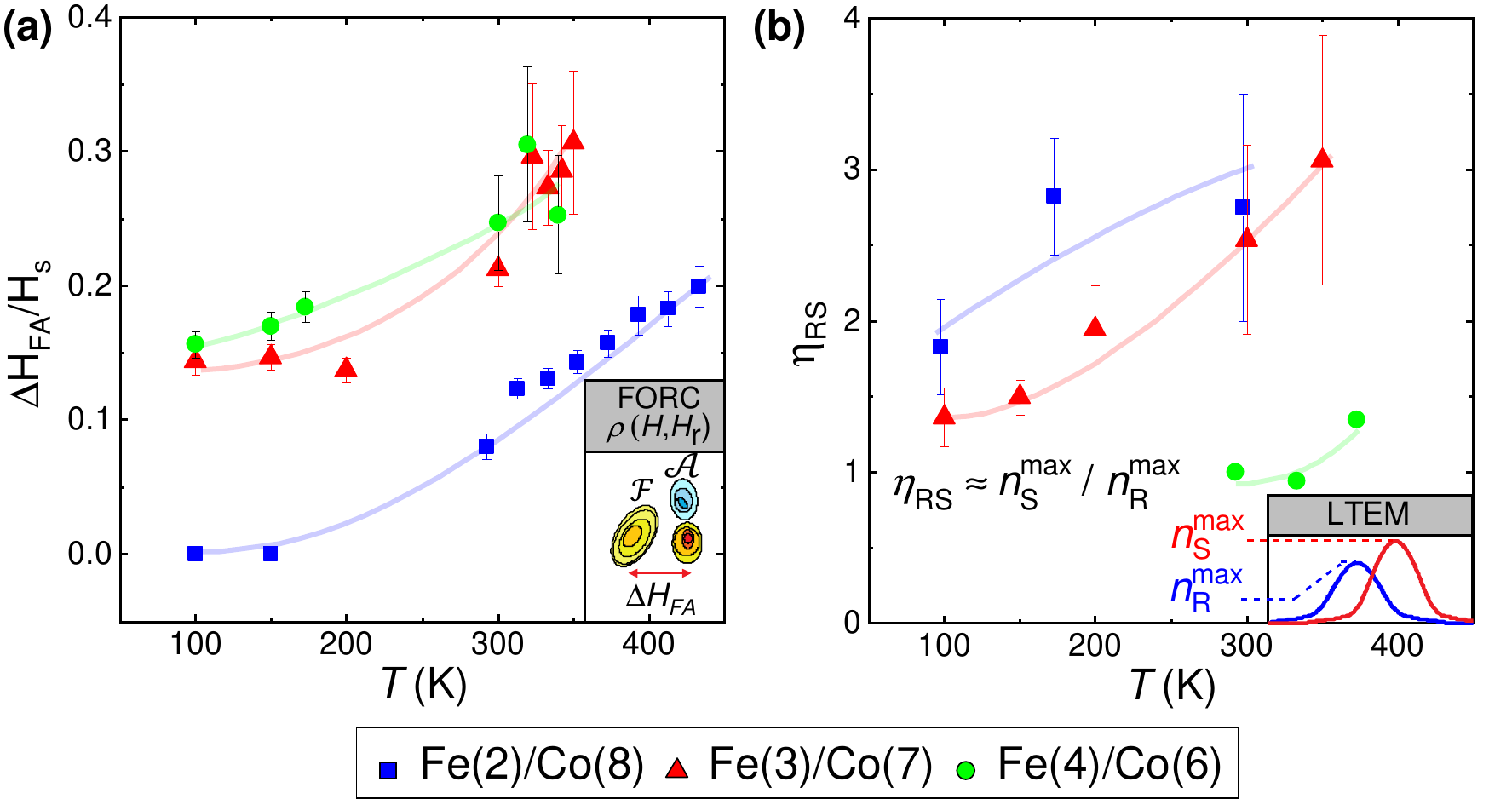}
\caption{\textbf{Sample-Wise $T$-Dependence of Stripe-to-Skyrmion Transition Characteristics.}
Compiled $T$-dependence across the three samples of
\textbf{(a)} FORC-measured separation of $\mathcal{F}$ and $\mathcal{A}$ peaks, $\Delta H_{\mathcal{F}\mathcal{A}}$, and 
\textbf{(b)} LTEM-measured ratio of maximal skyrmion to stripe densities, $\eta_{\rm RS} \equiv n_{\rm S}^{\rm max}$ / $n_{\rm R}^{\rm max}$.
The trends exhibit visible sample-wise variation c.f. manuscript Fig. 4(c-d).}
\label{fig:FORC-LTEM-TDep}
\end{figure}

In manuscript Fig. 4, we examined the thermodynamic variation of both the measured stripe-skyrmion transition characteristics across samples and temperatures --- (1) the separation, $\Delta H_{\mathcal{F}\mathcal{A}}$, of the $\mathcal{F}$ and $\mathcal{A}$ peaks measured by FORC magnetometry, and (2) the ratio of f maximal skyrmion to stripe densities, $\eta_{\rm RS} \equiv n_{\rm S}^{\rm max}$ / $n_{\rm R}^{\rm max}$.
While manuscript Fig. 4(c-d) showed the variation of $\Delta H_{\mathcal{F}\mathcal{A}}$ and $\eta_{\rm RS}$ with $\kappa_{est}$ (obtained from manuscript Fig. 4(b)), we present in \ref{fig:FORC-LTEM-TDep} the $T$-dependence of both these characteristics. 
Notably, we find that while both $\Delta H_{\mathcal{F}\mathcal{A}}$ (\ref{fig:FORC-LTEM-TDep}(a)) and $\eta_{\rm RS}$ (\ref{fig:FORC-LTEM-TDep}(b)) rise monotonically with $T$ for all three samples, the trends exhibit visible sample dependence. In contrast, manuscript Fig. 4(c-d) consistently show the collapse of sample- and $T$-dependent data onto a single curve. This provides further experimental support for the crucial role played by the thermodynamic parameter $\kappa$ in determining the nature of the stripe-to-skyrmion transition.

\noindent \begin{center}
{\small{}\rule[0.5ex]{0.6\columnwidth}{0.5pt}}{\small\par}
\par\end{center}

\end{document}